# Enhanced Superconductivity and Vortex Dynamics in One-Dimensional TaS$_2$ Nanowires

Mathew Pollard[1], Visakha Ho[1], Clarissa Wisner[2], Eric Bohannan[2], and Yew San Hor[1,2,a]

[1]Department of Physics, Missouri University of Science and Technology, Rolla 65409, USA

[2]Materials Research Center, Missouri University of Science and Technology, Rolla 65409, USA

a) Electronic mail: yhor@mst.edu

## Abstract

We report the synthesis of high-quality 2H-TaS$_2$ nanowires via a controlled two-step conversion process from TaS$_3$ precursors, achieving robust superconductivity with a transition temperature $T_c \approx 3.6$ K which is significantly higher than bulk 2H-TaS$_2$ ($T_c \approx 0.8$ K). Structural and compositional analyses confirm phase purity and preserved 1D morphology, while magnetotransport measurements reveal an enhanced upper critical field $\mu_0 H_{c2}$ (2 K) $\approx 5$ T, far exceeding the bulk value ($\mu_0 H_{c2}$ (0) $\approx 1.17$ T), attributed to dimensional confinement and suppression of charge density wave order. Magnetic characterization demonstrates complex vortex dynamics, including flux jumps and a second magnetization peak, indicative of strong pinning and crossover from elastic to plastic vortex regimes. These findings establish TaS$_2$ nanowires as a versatile platform for studying superconductivity in reduced dimensions and exploiting confinement-driven quantum phenomena for advanced applications.

Layered transition metal dichalcogenides (TMDs) have garnered significant attention due to their unique properties and potential for diverse applications. Their layered structure, which allows for exfoliation into atomically thin two-dimensional (2D) sheets, makes them attractive for electronic, optoelectronic, and energy storage applications [1-3]. They have attracted wide interest for their rich electronic phase diagrams, which include charge density waves (CDWs) and Mott insulating states [4-6]. The discovery of 2M-phase TMDs, for examples $MoS_2$, $WS_2$ has further expanded the landscape, revealing exotic properties such as superconductivity, valleytronics, and topological states, which are pivotal for next-generation electronics [7,8]. Concurrently, interface engineering in metal-TMD heterostructures such as $Pt/MoS_2$ has enabled precise control over charge transfer and catalytic activity, advancing applications in hydrogen evolution reactions and energy storage [9]. Meanwhile, engineered TMD composites (e.g., $MoS_2$/carbon-based materials) continue to set new benchmarks in supercapacitor energy density, driven by synergistic effects between enhanced conductivity and active sites, enabling highly effective energy storage solutions [10].

Among them, $TaS_2$ stands out due to its polymorphism and the fascinating interplay between competing collective quantum phenomena [11-13]. It crystallizes in several polytypes, notably the 1T, 2H, and 4Hb structures, each defined by distinct layer stacking and metal coordination. The 1T phase exhibits a sequence of CDW transitions as the temperature decreases, starting from an incommensurate CDW state and evolving into a nearly commensurate CDW phase at ~350 K, before finally transitioning to a fully commensurate CDW accompanied by a Mott insulating state below ~180 K [14-16]. In this low-temperature phase, the lattice forms a long-range ordered $\sqrt{13} \times \sqrt{13}$ superlattice of "Star of David" clusters, where twelve Ta atoms contract around a central Ta atom to create a localized electronic state [17]. External stimuli such as pressure, intercalation, or gating can suppress the Mott state and induce superconductivity, with reported $T_c$ values up to ~7 K [18-22].

The coexistence and competition of CDW and superconductivity in $TaS_2$ have made it a prime platform for studying correlated electron phenomena, including quantum phase transitions, dimensional crossover, and emergent non-equilibrium dynamics [23-25]. The 2H phase, which features trigonal prismatic coordination, exhibits superconductivity in the bulk with a superconducting transition temperature $T_c \approx 0.8$ K [26,27]. In the mixed-layer 4Hb polytype of $TaS_2$, $T_c$ is enhanced to approximately 2.2 K compared to the 0.8 K in the pure 2H phase [28]. This enhancement is attributed to a momentum-anisotropic increase in electron-phonon coupling and the suppression of CDW order, as revealed by angle-resolved photoemission spectroscopy (ARPES) measurements. Furthermore, $T_c$ in 2H-$TaS_2$ is enhanced when thinned down to a few atomic layers. Navarro-Moratalla *et al.* [28] observed that superconductivity persists down to the thinnest layer investigated (~3.5 nm), with a pronounced enhancement of $T_c$ from 0.5 K in the bulk to 2.2 K in thin layers. They attribute this phenomenon to an enhancement of the effective electron-phonon coupling constant resulting from reduced dimensionality. Specifically, Yang *et al.* [29] reported that in the monolayer limit, $T_c$ further increases to 3.4 K compared to 0.8 K in the bulk. This enhancement is ascribed to the suppression of CDW order and an increase in the density of states at the Fermi level due to reduced dimensionality.

Dimensional confinement thus plays a key role in enhancing superconductivity in $TaS_2$. Beyond 2D, further confinement into one-dimensional (1D) nanostructures such as nanobelts and nanowires, can raise $T_c$ even more. For example, $TaS_2$ nanobelts synthesized through a two-step process involving chemical vapor transport (CVT) of $TaS_3$ precursors followed by pyrolysis in vacuum exhibit $T_c$ up to 2.7 K [30]. Nanowires of $TaS_2$, synthesized directly from elements in a one-step reaction, have shown superconductivity up to 3.4 K [31]. However, making these $TaS_2$ nanostructures at scale and without defects is still challenging. In this work, we report a two-step synthesis method to produce high-quality $TaS_2$ nanowires with $T_c \approx 3.6$ K. Our process offers a scalable and reproducible route to 1D $2H$-$TaS_2$ nanostructures. This approach enables reliable production of high-quality nanowires with high aspect ratios, offering an exciting platform for studying superconductivity in confined geometries and investigating unique vortex behaviors such as avalanche dynamics.

The synthesis of $TaS_2$ superconducting nanowires can be achieved through a controlled conversion process [32] from $TaS_3$ nanowires. A critical aspect of this transformation is balancing two competing reactions: while sulfur (S) must be selectively removed from $TaS_3$ nanowires at elevated temperatures, excessive thermal exposure can degrade or oxidize the nanostructures. Direct annealing of $TaS_3$ nanowires in flowing inert gas above 300 °C often led to uncontrolled sulfur loss and the formation of tantalum oxide ($TaO_x$) phases, likely due to residual oxygen in the inert gas or outgassing from the quartz tube.

To overcome this, we employed a closed-system, vapor-phase reaction within a sealed quartz ampoule. The conversion is guided by the equilibrium reaction:

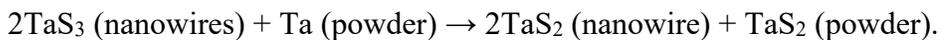

$2TaS_3$ (nanowires) + Ta (powder) → $2TaS_2$ (nanowire) + $TaS_2$ (powder).

In this setup as shown in Figure 1(a), excess Ta powder serves as a sulfur absorber, driving the reduction of $TaS_3$ to $TaS_2$ through a self-limiting reaction mechanism that automatically stops once the Ta powder is consumed, preserving the integrity of the nanostructures.

Precursor $TaS_3$ nanowires were first synthesized by mixing high-purity Ta (99.999%) and S (99.999%) powders in a Ta:S ratio of 1:3.03. The powders were thoroughly ground, sealed in an evacuated quartz ampoule (10 mm inner diameter, ~10 cm length), and heated to 750 °C at a rate of 1 °C/min. The temperature was held for 48 h, followed by controlled cooling to room temperature at 2 °C/min. Powder x-ray diffraction (XRD, monochromatic CuKα, λ=1.540598 Å) confirmed the formation of the monoclinic TaS$_3$ phase (Figure 2(a)).

The as-grown TaS$_3$ nanowires were then sealed in a quartz ampoule with Ta powder, placed separately to avoid direct mixing as shown in Figure 1(a). The ampoule was evacuated to ~$10^{-6}$ torr and heated to 590 °C for 12 h. It was then cooled to room temperature at a controlled rate of 2 °C/min to minimize thermal shock and maintain nanowire morphology. This process yielded TaS$_2$ nanowires and TaS$_2$ powder in separate regions of the ampoule, confirmed by SEM (Figures 1(c)) and XRD (Figure 2(b)). The optimal conversion temperature 590 °C was determined to prevent overheating and ensure the preservation of the nanostructures. The resulting nanowires have cross-sectional widths of 80-700 nm and lengths up to a millimeter. SEM and energy-dispersive X-ray spectroscopy (EDS) confirmed that the nanowires retained their original 1D morphology and were phase-pure 2H-TaS$_2$.

Low-temperature transport measurements were performed on bundles of TaS$_2$ and TaS$_3$ nanowires using a Quantum Design PPMS, with two-probe contacts made from room-temperature-cured silver paste. Figure 3(a) shows the temperature dependence of the resistance at zero magnetic field. The TaS$_3$ nanowires display semiconducting behavior with no signs of superconductivity down to 2 K, consistent with their quasi-1D CDW-dominated characteristics. In contrast, the converted TaS$_2$ nanowires exhibit a clear superconducting transition at $T_c \approx 3.6$ K (inset of Figure 3(a)), significantly higher than the bulk $T_c \approx 0.8$ K of 2H-TaS$_2$, reflecting the effects of confinement and enhanced electron-phonon coupling. Notably, no CDW transition is observed in the TaS$_2$ nanowire resistance data. Figure 3(b) shows the magnetic field dependence of resistance for the TaS$_2$ nanowires at various temperatures below $T_c$. The superconducting transition shifts to lower fields with increasing temperature, and the corresponding upper critical field ($H_{c2}$) values are extracted from these curves. The inset of Figure 3(b) plots the temperature dependence of $H_{c2}$, revealing a clear linear decrease with increasing temperature. The observed enhancement of $\mu_0 H_{c2} \approx 5$ T at 2 K in these 1D nanowires compared to that of bulk values ($\mu_0 H_{c2}(0) \approx 1.17$ T reported by Liu *et al.* [33]) is consistent with the confinement-induced increase in the superconducting condensation energy and the reduced free-energy density of the superconducting state. This behavior can be described by the relationship $H_{c2}^*/H_{c2} \propto 8\lambda/d$ where $H_{c2}^*$ is the enhanced upper critical field in the confined geometry (such as in nanowires), $H_{c2}$ is the upper critical field in the bulk material, $\lambda$ is the London penetration depth, and $d$ is the wire diameter. The enhancement factor $H_{c2}^*/H_{c2}$ quantifies how much the upper critical field increases in reduced dimensions compared to bulk. This effect has been previously confirmed in other 1D superconducting systems, such as Pb nanowires [34]. In our TaS$_2$ nanowires, this enhancement is attributed to the suppression of CDW order and an increase in the density of states at the Fermi level due to reduced dimensionality. These results demonstrate that reducing TaS$_2$ to 1D nanowires not only enhances the superconducting transition temperature but also significantly boosts the upper critical field $H_{c2}$.

Figure 4(a) presents the temperature dependence of the magnetization of TaS$_2$ and TaS$_3$ nanowires measured by AC susceptibility. The real part of the AC magnetic susceptibility ($M'$) for TaS$_2$ nanowires shows a sharp diamagnetic onset at approximately $T_c \approx 3.6$ K, consistent with the superconducting transition observed in transport measurements. In contrast, the TaS$_3$ nanowires exhibit no such transition down to 2 K, indicating the absence of superconductivity and consistent with their quasi-one-dimensional charge-density-wave–dominated behavior. The inset of Fig. 4(a) displays the DC magnetization ($M_{DC}$) of the TaS2 nanowires measured under an applied field of 10 Oe. The zero-field-cooled (ZFC) curve shows a pronounced diamagnetic signal below $T_c$ upon warming from 2 K to 5 K, indicating strong shielding due to superconductivity, then the field-cooled (FC) data, collected while cooling from 5 K to 2 K, exhibits a less negative magnetization, as expected due to flux trapping. Arrows in the inset mark the direction of temperature sweep for both ZFC and FC conditions. Together, these AC and DC magnetization results confirm the bulk superconductivity in the $TaS_2$ nanowire bundles and the absence of superconducting response in $TaS_3$.

Figure 4(b) presents isothermal magnetization $M_{DC}$ as a function of applied magnetic field (*MH* loop) measured at 2 K. The data exhibit pronounced flux jumps and a second peak (SP) or "fishtail" effect which are characteristics commonly associated with strong vortex pinning in type-II superconductors. These flux jumps, also known as magnetic instabilities, result from abrupt vortex motion and are a critical aspect of both fundamental vortex dynamics and practical applications of superconductors. In type-II superconductors, above the lower critical field ($H_{c1}$), the magnetic field penetrates the bulk as quantized vortices, which can form a triangular lattice in ideal conditions. However, disorder and pinning centers disrupt this order, leading to a critical state. Under thermal fluctuations or rapid field changes, this state can become unstable, causing sudden vortex motion that appears as discontinuities as flux jumps in *MH* loop measurements. These jumps are known to reduce the critical current density $J_c$ and can drive the superconductor closer to the normal state, impacting applications [35,36]. This phenomenon has been extensively studied in both low-temperature superconductors (e.g., Nb-based systems [37]) and high-temperature superconductors (e.g., cuprates and $MgB_2$ [38]), where similar second-peak features and flux jumps have been observed. In $TaS_2$ nanowires, these effects highlight the presence of collective vortex pinning and suggest that the second peak in the magnetization likely originates from a crossover in the vortex pinning mechanism from elastic to plastic regimes. The data also highlight the enhanced magnetic stability of these nanowires, with the field of the first flux jump occurring at higher-than-predicted values—potentially due to significant flux creep.

In summary, we have demonstrated the successful conversion of $TaS_3$ nanowires to high-quality $2H\text{-}TaS_2$ nanowires via the two-step synthesis method that maintains the one-dimensional morphology and phase purity of the resulting structures. The resulting $TaS_2$ nanowires exhibit a superconducting transition at $T_c \approx 3.6$ K, significantly higher than the bulk value, confirming robust bulk superconductivity in these confined geometries. Low-temperature magneto transport and magnetization measurements reveal an enhanced upper critical field ($H_{c2} \approx 5$ T at 2 K), consistent with theoretical predictions for 1D superconductors and attributed to confinement-induced enhancement of the superconducting condensation energy and the suppression of charge density wave order. Additionally, pronounced flux jumps and a second peak in the magnetization loops highlight the presence of strong vortex pinning and complex vortex dynamics in these nanowires. Our results demonstrate that dimensional confinement in $TaS_2$ not only enhances the superconducting transition temperature but also significantly boosts the upper critical field and alters the vortex behavior, offering an exciting platform for probing low-dimensional superconductivity, phase fluctuations, and collective vortex phenomena.

**Acknowledgments**

One of the authors Y. S. H. acknowledges support from the Materials Research Center.

# References

[1] Q. H. Wang, K. Kalantar-Zadeh, A. Kis, J. N. Coleman, and M. S. Strano, *Nature Nanotech.* **7**, 699 (2012).

[2] M. Chhowalla, H. S. Shin, G. Eda, L. J. Li, K. P. Loh, and H. Zhang, *Nature Chem.* **5**, 263 (2013).

[3] S. Manzeli, D. Ovchinnikov, D. Pasquier, O. V. Yazyev, and A. Kis, *Nat. Rev. Mater.* **2**, 17033 (2017).

[4] Qiang Gao, Haiyang Chen, Wen-shin Lu, Yang-hao Chan, Zhenhua Chen, Yaobo Huang, Zhengtai Liu, and Peng Chen, *Nat. Commun.* **16**, 3784 (2025).

[5] X. Zhang. S. Yan, and G. Li, *Phys. Rev. B* **110**, 235137 (2024).

[6] C. J. Sayers, H. Hedayat, A. Ceraso, F. Museur, M. Cattelan, L. Hart, L. Farrar, S. Dal Conte, G. Cerullo, C. Dallera, E. Como, and E. Carpene, *Phys. Rev. B* **102**, 161105(R) (2020).

[7] J. M. Lu, O. Zheliuk, I. Leermakers, N. F. Q. Yuan, U. Zeitler, K. T. Law, and J. T. Ye, *Science* **350**, 1353 (2015).

[8] Y. Fang, J. Pan, D. Q. Zhang, D. Wang, H. T. Hirose, T. Terashima, S. Uji, Y. Yuan, W. Li, Z. Tian, J. Xue, Y. Ma, W. Zhao, Q. Xue, G. Mu, and H. J. Zhang, *Adv. Mater.* **31**, 1901942 (2019).

[9] A. Shan, X. Teng, Y. Zhang, P. Zhang, Y. Xu, C. Liu, H. Li, H. Ye, and R. Wang, *Nano Energy* **94**, 106913 (2022).

[10] M. Szkoda, A. Ilnicka, K. Trzciński, Z. Zarach, D. Roda, and A. P. Nowak, *Sci. Rep.* **14**, 26128 (2024).

[11] B. Sipos, A. F. Kusmartseva, A. Akrap, H. Berger, L. Forró, and E. Tutiš, *Nat. Mater.* **7**, 960 (2008).

[12] Y. Yu, F. Yang, X. F. Lu, Y. J. Yan, Y. H. Cho, L. Ma, X. Niu, S. Kim, Y. W. Son, D. Feng, S. Li, S. W. Cheong, X. H. Chen, and Y. Zhang, *Nature Nanotech.* **10**, 270 (2015).

[13] T. Ritschel, J. Trinckauf, K. Koepernik, B. Büchner, M. v. Zimmermann, H. Berger, Y. I. Joe, P. Abbamonte, and J. Geck, *Nature Phys.* **11**, 328 (2015).

[14] P. Fazekas and E. Tosatti, *Philos. Mag. B* **39**, 229 (1979).

[15] Y. D. Wang, W. L. Yao, Z. M. Xin et al., *Nat. Commun.* **11**, 4215 (2020).

[16] Q. Gao, H. Chen, W. Lu, Y. Chan, Z. Chen, Y. Huang, Z. Liu, and P. Chen, *Nat. Commun.* **16**, 3784 (2025).

[17] Y. Fei, Z. Wu, W. Zhang, and Y. Yin, *AAPPS Bull.* **32**, 20 (2022).

[18] S. Sajan, H. Guo, T. Agarwal, I. Sánchez-Ramírez, C. Patra, M. G. Vergniory, F. de Juan, R. P. Singh, and M. M. Ugeda, *Nano Lett.* **25**, 6654 (2025).

[19] B. Sipos, A. F. Kusmartseva, A. Akrap, H. Berger, L. Forró, and E. Tutis, *Nat Mater.* **7**, 960 (2008).

[20] Q. Dong, Q. Li, S. Li, X. Shi, S. Niu, S. Liu, R. Liu, B. Liu, X. Luo, J. Si, W. Lu, N. Hao, Y. Sun, and B. Liu, *npj Quantum Mater*. **6**, 20 (2021).

[21] W. Zhang, D. Ding, J. Gao et al., *Nano Res.* **15**, 4327 (2022).

[22] C. Boix-Constant, S. Mañas-Valero, R. Córdoba, J. J. Baldoví, Á. Rubio, and E. Coronado, *ACS Nano* **15**, 11898 (2021).

[23] M. Kratochvilova, A.D. Hillier, A. R. Wildes *et al*., *npj Quantum Mater*. **2**, 42 (2017).

[24] S. Wang, Y. Han, S. Sun, S. Wang *et al*., *Phys. Rev. Lett.* **133**, 056001 (2024).

[25] Y. Kvashnin, D. VanGennep, M. Mito, S. A. Medvedev, R. Thiyagarajan, O. Karis, A. N. Vasiliev, O. Eriksson, and M. Abdel-Hafiez, *Phys. Rev. Lett.* **125**, 186401 (2020).

[26] J. M. E. Harper, T. H. Geballe, and F. J. DiSalvo, *Phys. Rev. B* **15**, 2943 (1977).

[27] E. Navarro-Moratalla, J. O. Island, S. Mañas-Valero, E. Pinilla-Cienfuegos, A. Castellanos-Gomez, J. Quereda, *et al*., *Nat. Commun.* **7**, 11043 (2016).

[28] W. R. Pudelko, H. Liu, F. Petocchi, H. Li, E. B. Guedes, J. Küspert, K. von Arx, Q. Wang, R. C. Wagner *et al*., *Phys. Rev. Materials* **8**, 104802 (2024).

[29] Y. Yang, S. Fang, V. Fatemi, J. Ruhman *et al*., *Phys. Rev. B* **98**, 035203 (2018).

[30] X. Wu, Y. Tao, Y. Hu, Y. Song, Z. Hu, J. Zhu, and L. Dong, *Nanotechnology* **17**, 201 (2006).

[31] C. W. Dunnill, H. K. Edwards, P. D. Brown, D. H. Gregory, *Angewandte Chemie* **45**, 7060 (2006).

[32] Y. S. Hor, U. Welp, Y. Ito, Z. L. Xiao, U. Patel, J. F. Mitchell, W. K. Kwok, and G. W. Crabtree, *Appl. Phys. Lett.* **87**, 142506 (2005).

[33] H. Liu, S. Huangfu, H. Lin, X. Zhang, and A. Schilling, *J. Mater. Chem. C* **11**, 3553 (2023).

[34] M. He, C. H. Wong, P. L. Tse, Y. Zheng, H. Zhang, F. L. Y. Lam, P. Sheng, X. Hu, and R. Lortz, *ACS Nano* **7**, 4187 (2013).

[35] R. G. Mints, and A. L. Rakhmanov, *Rev. Mod. Phys.* **53**, 551 (1981).

[36] A. Altshuler, and T. H. Johansen, *Rev. Mod. Phys.* **76**, 471 (2004).

[37] A. Nabialek, M. Niewczas, H. Dabkowska, A. Dabkowski, J. P. Castellan, and B. D. Gaulin, *Phys. Rev. B* **67**, 024518 (2003).

[38] C. Romero-Salazar, F. Morales, R. Escudero, A. Duran, and O. A. Hernandez-Flores, *Phys. Rev. B* **76**, 104521 (2007).

**Figure Captions**

**Figure 1.** Synthesis and morphological characterization of $TaS_3$ and $TaS_2$ nanowires. (a) A schematic of the sealed quartz ampoule setup used to convert $TaS_3$ to $TaS_2$ nanowires in vacuum conditions. $TaS_3$ nanowires and Ta powder were positioned at opposite ends of the quartz tube to prevent direct mixing during the vapor-phase reaction. (b) SEM images of the as-synthesized $TaS_3$ nanowires, revealing a dense network of long, flexible nanowires with diameters down to the nanoscale, as highlighted in the high-magnification inset. (c) SEM images of the resulting $TaS_2$ nanowires after the conversion process, demonstrating that the one-dimensional morphology is well retained, as confirmed by the higher-magnification inset.

**Figure 2.** XRD patterns of (a) $TaS_3$ nanowires and (b) $TaS_2$ nanowires. The $TaS_3$ nanowires display sharp diffraction peaks that match the monoclinic phase, confirming high crystallinity and phase purity. The $TaS_2$ nanowires exhibit dominant peaks corresponding to the hexagonal $2H$-$TaS_2$ phase, with an intense (002) reflection indicating preferred orientation along the c-axis. Several weak peaks marked by asterisks (*) might be due to minor secondary phases or slight residual precursor traces.

**Figure 3.** (a) Temperature dependence of resistance ($R$) at zero magnetic field ($H = 0$) for bundles of $TaS_2$ nanowires (red) and $TaS_3$ nanowires (black). The $TaS_2$ nanowires exhibit a clear superconducting transition at $T_c \approx 3.6$ K, as shown in the inset. (b) Magnetic field dependence of resistance ($R$) for $TaS_2$ nanowires at various temperatures below $T_c$, showing the suppression of superconductivity with increasing field. The inset plots the upper critical field ($H_{c2}$) as a function of temperature ($T$), revealing a linear decrease with increasing temperature.

**Figure 4.** (a) Temperature dependence of the magnetization of $TaS_2$ (red) and $TaS_3$ (green) nanowires. The main panel shows the real part of the AC susceptibility $M'$, while the inset presents

the DC magnetization ($M_{DC}$) measured under zero-field-cooled (ZFC) and field-cooled (FC) conditions at an applied field of 10 Oe. Arrows in the inset indicate the direction of the measurement. A sharp diamagnetic onset at $T_c \approx 3.6$ K in the TaS$_2$ nanowires confirms the superconducting transition. (b) Isothermal DC magnetization ($M_{DC}$) as a function of applied magnetic field for TaS$_2$ nanowires measured at 2 K. The magnetization hysteresis loop displays pronounced flux jumps and a second peak ("fishtail" effect), indicating strong vortex pinning and magnetic instabilities in low-field regions. Arrows in the graph denote the direction of the data acquisition.

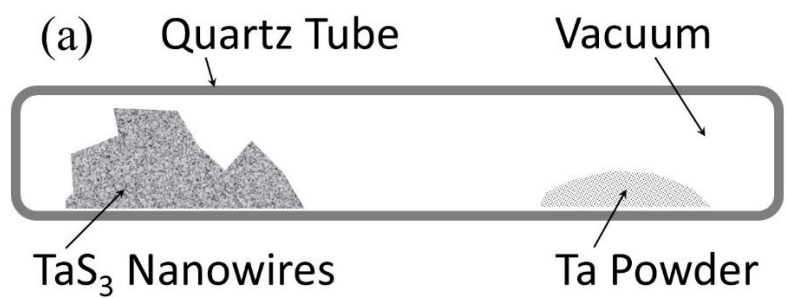
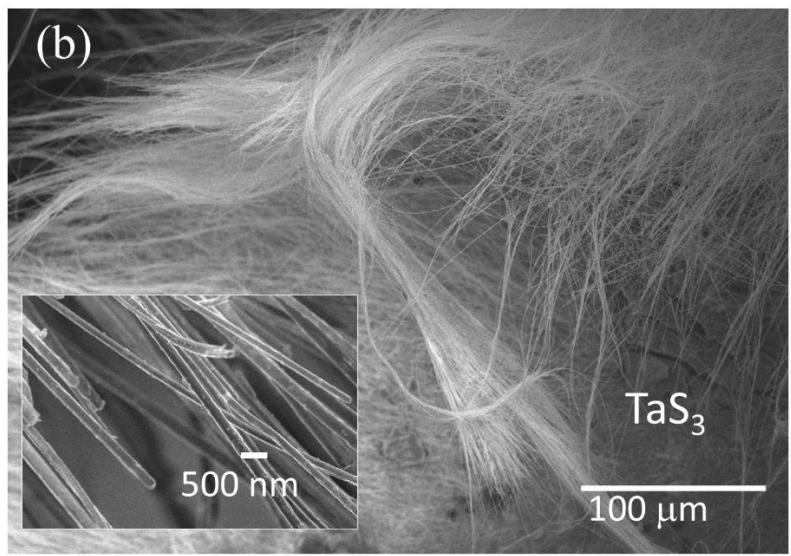
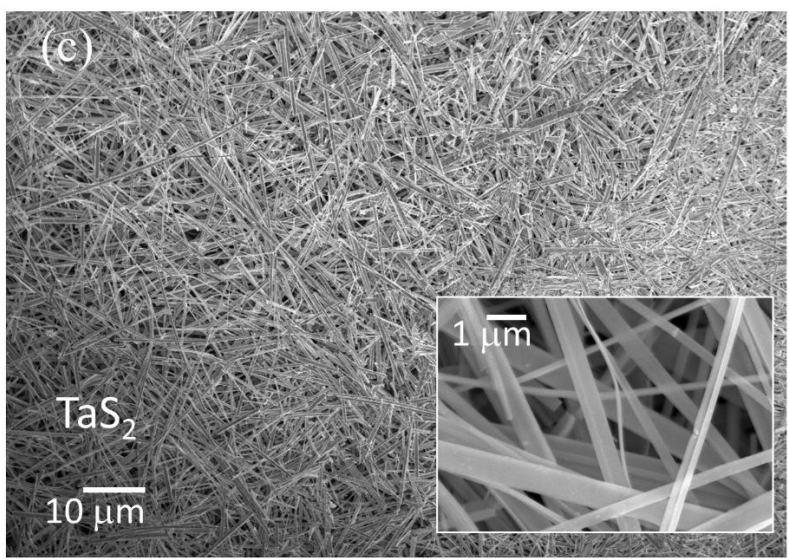

Figure 1

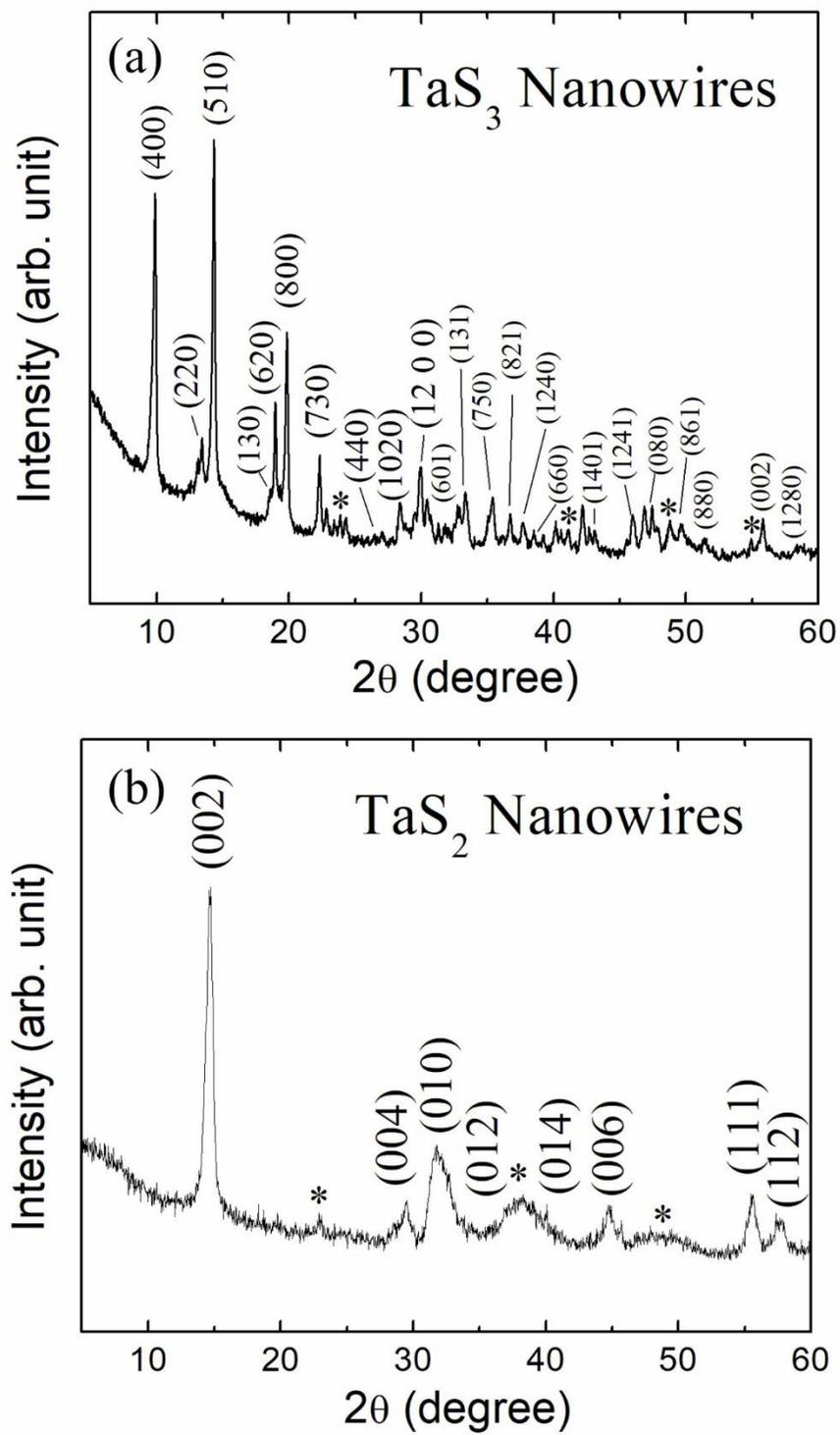

Figure 2

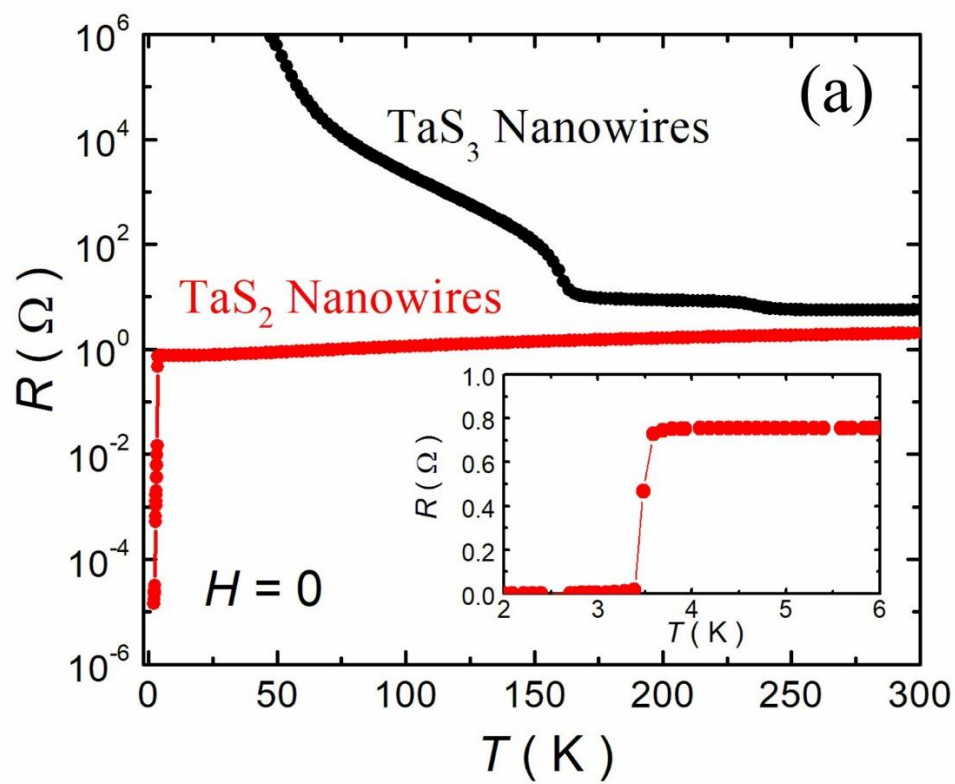
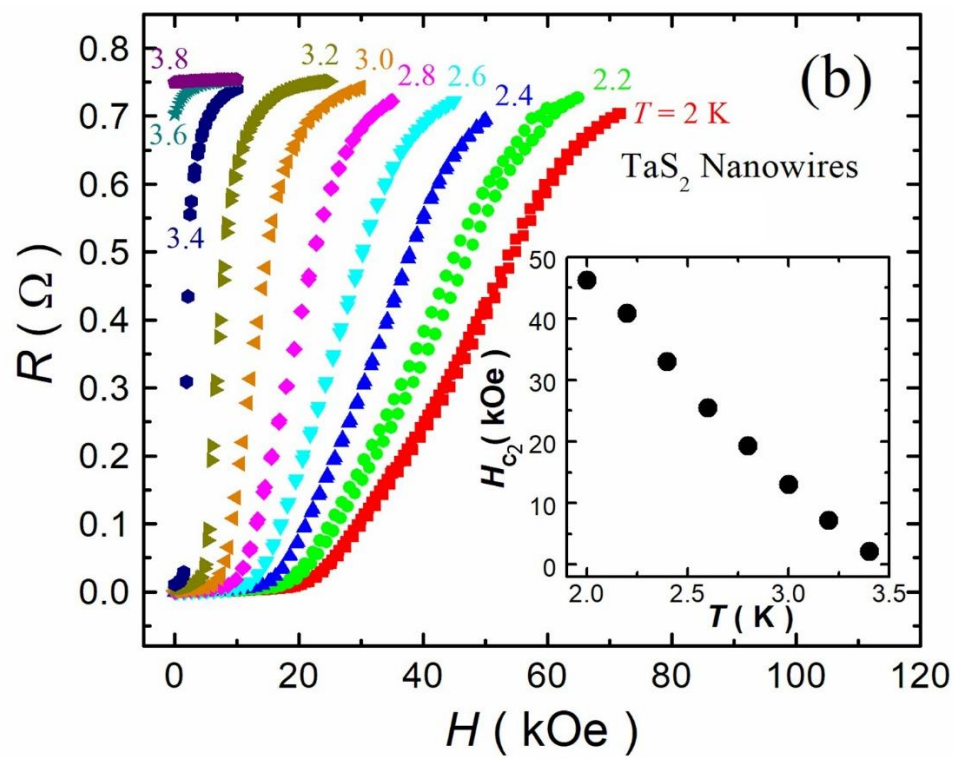

Figure 3

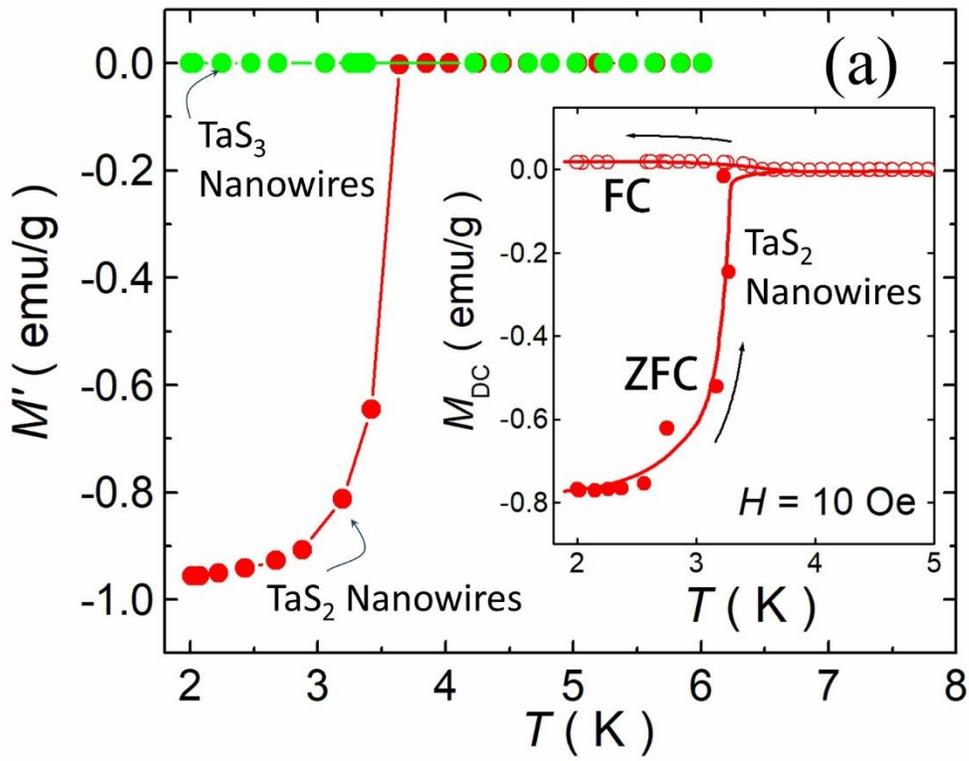

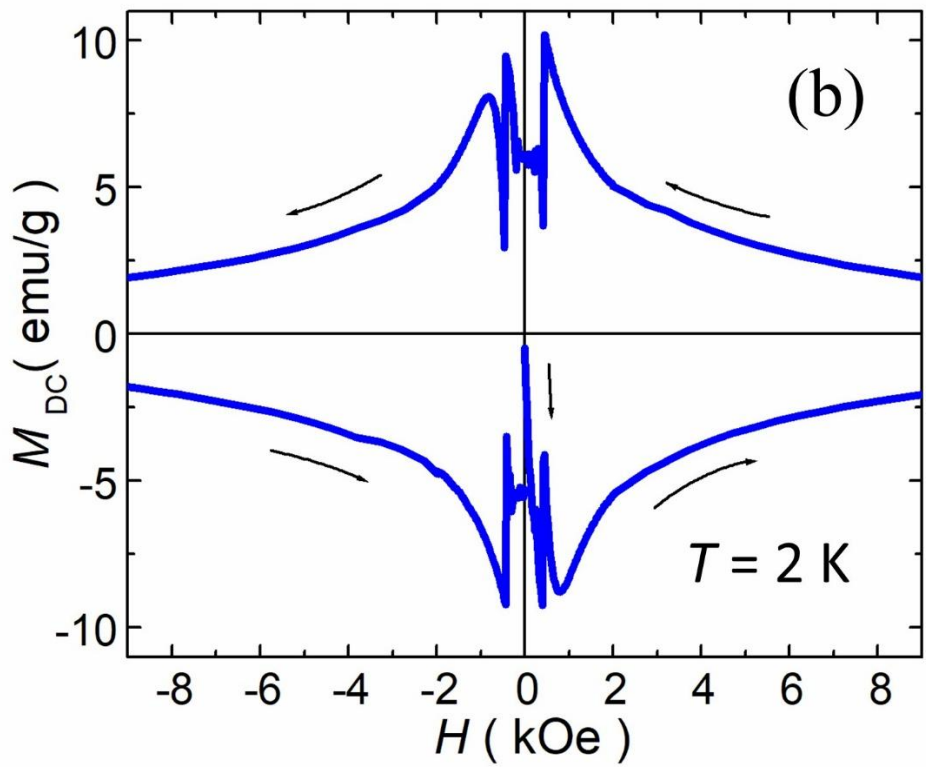

Figure 4